\documentclass[prl,preprint,showpacs,superscriptaddress]{revtex4-1}

\usepackage{graphicx}
\usepackage{dcolumn}
\usepackage{bm}

\begin{document}


\title{Dirac cone with helical spin polarization in ultrathin $\alpha$-Sn(001) films}

\author{Yoshiyuki Ohtsubo}
\email{yoshiyuki.ohtsubo@synchrotron-soleil.fr}
\affiliation{Synchrotron SOLEIL, Saint-Aubin-BP 48, F-91192 Gif sur Yvette, France}
\author{Patrick Le F\`evre}
\author{Fran\c{c}ois Bertran}
\affiliation{Synchrotron SOLEIL, Saint-Aubin-BP 48, F-91192 Gif sur Yvette, France}
\author{Amina Taleb-Ibrahimi}
\email{amina.taleb@synchrotron-soleil.fr}
\affiliation{Synchrotron SOLEIL, Saint-Aubin-BP 48, F-91192 Gif sur Yvette, France}
\affiliation{UR1/CNRS–Synchrotron SOLEIL, Saint-Aubin, F-91192 Gif sur Yvette,
France}

\date{\today}

\begin{abstract}
Spin-split two-dimensional electronic states have been observed on ultrathin Sn(001) films grown on InSb(001) substrates.
Angle-resolved photoelectron spectroscopy (ARPES) performed on these films revealed Dirac-cone-like linear dispersion around the $\bar{\Gamma}$ point of surface Brillouin zone, suggesting nearly massless electrons belonging to 2D surface states.
The states disperse across a bandgap between bulk-like quantum well states in the films.
Moreover, both circular dichroism of ARPES and spin-resolved ARPES studies show helical spin polarization of the Dirac-cone-like surface states, suggesting a topologically protected character as in a bulk topological insulator (TI).
These results indicate that a quasi-3D TI phase can be realized in ultrathin films of zero-gap semiconductors.
\end{abstract}

\maketitle

Topological insulators (TIs) are emerging as a new state of quantum matter with a bulk band gap and odd number of relativistic Dirac fermions, characterized by spin-polarized massless Dirac-cone (DC) dispersion of the edge/surface states \cite{Murakami03, Konig07, Fu07, Kane10, Qi11}.
The unique properties of surface electrons of TIs are an encouraging playground to realize new electronic phenomena, such as 
quantum spin-Hall effect \cite{Murakami03, Bernevig06, Konig07} and dissipation-less electron/spin transport \cite{Murakami03, Roth09, Xiu11}. 

For an application of the exotic surface states of TI to electronic/spintronic devices,
one of the most promising candidates are epitaxially-grown films of TIs, such as Bi$_2$Se$_3$ on SiC(0001) \cite{Zhang10} and Si(111) \cite{Zhang09, Sakamoto10},  HgTe on CdTe(001) \cite{Brune11, Yao13, Olivier13}, and $\alpha$-Sn on InSb(001) \cite{Barfuss13}.
They grow on semiconductor substrates and hence they could be combined with ordinal semiconductor electronic devices without any significant modification.
Furthermore, the epitaxial growth itself can manipulate the electronic structure of grown TI films.
Structural strain introduced from a lattice mismatch with the substrate opens a bandgap in bulk bands of HgTe \cite{Brune11, Yao13, Olivier13} or $\alpha$-Sn \cite{Barfuss13} films for thicknesses between 0.1 to 1 $\mu$m. 
While the size of the bandgap is only a few tens of meV, it is very important for these films as a TI; neither $\alpha$-Sn nor HgTe without any strain are TIs but zero-gap semimetal.
Quantum-size effect (QSE) \cite{Shik} from film thicknesses changes the electronic structure of TI even further.
Once the electrons are confined in very thin thickness, the bulk-band dispersion perpendicular to the film is no longer continuous but discrete, forming quantum-well (QW) states.
Such quantization of bulk bands enhances the bulk bandgap, as observed on the Bi$_2$Se$_3$ film \cite{Zhang10} and the layered compound (PbSe)$_5$(Bi$_2$Se$_3$)$_{3m}$ \cite{Nakayama12}.
The thickness of ultrathin films also influences surface DC to open a gap at the Dirac point, because of the hybridization between the top and bottom DCs \cite{Zhang10, Shan10}.
As mentioned above, ultrathin films provide a fertile ground to manipulate the electronic structure of materials concerning TI.

Zero-gap semimetal $\alpha$-Sn (gray-Sn) is a good candidate to examine the influence of QSE onto the electronic structure of TI.
Once finite bandgap is introduced in $\alpha$-Sn, it is predicted to become a 3D TI from theoretical calculations of the bulk bands \cite{Pollak70, Fu07}.
Although the $\alpha$ phase of bulk Sn is not stable at room temperature (RT), the epitaxial Sn films grown on lattice-matched substrates, such as InSb(001) \cite{Farrow81, Magnano02, Betti02}, InSb(111) \cite{Hernandez85, Kasukabe88, Kondo04}, and CdTe(001) \cite{Farrow81, Tang87}, show the stable $\alpha$ phase even above RT.
This is because the lattice-matched substrate stabilizes the $\alpha$ phase of the epitaxial Sn film.
While the growth behavior of such $\alpha$-Sn films is well known, the surface band structure, that would hold DC with non-zero bandgap from QSE, has never been studied so far.

In this letter, we report the surface-state evolution of $\alpha$-Sn(001) films grown on InSb(001) with various thickness of the films.
In a certain range of thickness, the surface state showed DC-like dispersion, measured with angle-resolved photoelectron spectroscopy (ARPES).
Both spin-resolved ARPES and circular dichroism of ARPES showed helical spin polarization of the DC-like surface state.
With smaller thicknesses, we also observed the gap opening onto the DC-like surface states.
Based on these results, we demonstrated that QSE can open a bulk bandgap in zero-gap semimetal and realize an ultrathin quasi-3D TI.

We grew the $\alpha$-Sn(001) films on InSb(001) substrates covered with 1 ML of Bi (1 ML is defined as the atom density of bulk-truncated InSb(001)).
With this procedure, Bi segregates at the surface during the Sn growth and forms the topmost atomic plane of the sample.
Our Bi/Sn(001) films showed low-background and sharp-spot low-energy electron diffraction (LEED) pattern as shown in Fig. 1 (a), indicating the growth of a well-ordered Bi/Sn(001) film.
Based on the double-domain (2$\times$1) periodicity in the LEED pattern, Bi atoms possibly form a dimer row, terminating the dangling bonds on the surface.
The detailed procedure of the sample growth, its characterization, and the role of the surface Bi are shown in the supplemental material (SM) \cite{SM}. 

ARPES measurements were performed at the CASSIOPEE beamline (SOLEIL, France) with an energy resolution of 20 meV, using linearly and left-/right-circularly polarized lights. 
The photon-incident plane is ($\bar{1}$10) and the electric field of linearly-polarized photons lies in the incident plane.
Figure 1 (b) and (c) are the Fermi contour and the band dispersion along [110] on a 30 ML Bi/Sn(001) film, respectively. 
There are no metallic electronic states crossing the Fermi level ($E_{\rm F}$) except at $\bar{\Gamma}$.
This state does not exhibit the (2$\times$1) surface periodicity, suggesting that it originates from the subsurface Sn layers.

Figure 2 shows a series of ARPES intensity plots of the Bi/Sn(001) films from 12 to 34 ML taken along [$\bar{1}$10] (a-e) and [110] (f-j) with linearly-polarized photons.
At 12 ML, there are two bands $S_1$ and $S_2$ dispersing upwards and downwards from $\bar{\Gamma}$, respectively, as shown in fig. 2(a, f).
$S_2$ disperses almost linearly, showing quite a small effective mass.
$S_1$ and $S_2$ show no energy shifts with different photon energies, indicating a 2D character.
There is a finite gap of 150 meV between $S_1$ and $S_2$. 
$S_2$ shifts upwards at 20 and 24 ML and $S_1$ is no longer observable below $E_{\rm F}$ for these thicknesses.
A way to observe electronic states above $E_{\rm F}$ is presented in the following part.
The dispersions of $S_2$ up to 24 ML are isotropic along [110] and [$\bar{1}$10].
On the thicker films (30 and 34 ML), $S_2$ shifts downwards and $S_1$ appears again along [$\bar{1}$10];
the gap between $S_1$ and $S_2$ is now of 200 meV.
However, the measurement along [110] shows another surface state $S_2'$ (see fig. 2(i, j)), dispersing between $S_2$ and $S_1$, which degenerates with $S_1$ at $\bar{\Gamma}$.
Hence, films thicker than 30 ML support metallic surface states which disperse across $E_{\rm F}$ continuously: this is a characteristic of topologically protected surface states of TIs.
In addition, we observe another feature along [110] whose upper edge is indicated by a dotted line in Fig. 2(j).
It does not show any obvious peak, suggesting its origin from bulk-like QW bands.
Since the surface has two domains with 90$^\circ$ rotation, as shown by the LEED pattern, the dispersion of the initial states should be isotropic along these two directions. 
The differences between them are due to the parities of the surface states with respect to the measurement planes: (110) for (a-e) and ($\bar{1}$10) for (f-j).
Detailed discussion is provided in SM \cite{SM}.

In order to obtain the surface-state band dispersion above $E_{\rm F}$, we divided the ARPES spectra at 24 ML by Fermi-Dirac distribution function convolved with the instrumental resolution to take into account thermally populated electrons there.
To increase the populations of thermally-excited electrons, the measurement was done at 450 K.
The result, Fig. 3 (a), shows a linear DC without any gap, with a Dirac point at $\sim$20 meV above $E_{\rm F}$.
The velocity of electrons is 7.3$\times$10$^5$ m/s, uniform from 0.6 eV to -0.15 eV.
At Dirac point, it is much larger than those of typical 3D TIs, such as Bi$_2$Se$_3$ (2.9$\times$10$^5$ m/s) and TlBiSe$_2$ (3.9$\times$10$^5$ m/s) \cite{Kuroda10}, and close to that of graphene (1$\times$10$^6$ m/s) \cite{Berger06}.

One of the most salient character of DC on TI is the helical spin polarization.
In other words, the electrons belonging to such states are spin-polarized toward the direction perpendicular to both the wave vector $k_{\parallel}$ and the surface normal.
In order to evaluate the polarization of the surface state on the Bi/Sn(001) films, we have measured both circular dichroism of ARPES and spin-resolved ARPES.
As depicted schematically in Fig. 3(b), the incident circularly polarized photons were in the ($\bar{1}$10) plane in our experimental geometry, and hence the helicity of the photons should probe the spin polarization along [110] or [001] \cite{Wang11}.
Figure 3(c) is the circular dichroism map measured along [$\bar{1}$10].
To also observe the upper part of DC,
we measured dichroism of ARPES in the 30 ML film, where both $S_1$ and $S_2$ are below $E_{\rm F}$.
They both show clear dichroic effect.
On $S_1$, it is positive (negative) for $k_y >$0 ($k_y <$0); it is the opposite for $S_2$.
Such dichroic behavior is consistent with the helical spin polarizations of the non-trivial surface states on TI.
In addition, there is another feature dispersing downwards from 0.2 eV to 0.6 eV with $|k_y| >$0.15 \AA$^{-1}$.
It overlaps the peakless ARPES feature observed in Fig. 2 (j).

Figures 3(d, e) are spin-resolved ARPES spectra measured at $k_y$ = 0.07 and -0.07 \AA$^{-1}$, respectively, with linearly-polarized photons at h$\nu$ = 19 eV.
The overall energy resolution was set to 120 meV and the angular acceptance of the detector was $\pm$1.8$^\circ$ ($\pm$0.06 \AA$^{-1}$ with 19 eV photons).
As shown in the spin-resolved spectra, $S_1$ is spin polarized towards [$\bar{1}\bar{1}$0] ([110]) at positive (negative) value of $k_y$ and those for $S_2$ are the opposite.
They correspond to clockwise (CW) helicity for the lower cone $S_2$ and counter CW (CCW) for the upper cone $S_1$.
Simultaneously, we also measured the spin polarization along the surface normal.
It shows almost negligible polarizations, indicating that the spin polarization of the DC on the Sn(001) films lies almost completely in the in-plane direction.
It could be due to the absence of the $C_3$ symmetry on the (001) surface \cite{Fu09}.

To obtain further insight into the electronic structure of the Bi/Sn(001) films, we performed density-functional-theory (DFT) calculations. We used the ``augmented plane wave + local orbitals" method implemented in the WIEN2k code \cite{Blaha01} taking spin-orbit interaction into account. 
We adopted the modified Becke and Johnson potential together with the exchange-correlation potential constructed by using the local density approximation \cite{Becke06, Tran09}. 
The film was modeled by a symmetric slab of 32 Sn layers with the surface covered with (2$\times$1) dimers of Bi, the structure of which was energetically optimized down to the 12th Sn layer.

Figure 4(a) shows the calculated states along [110] and [$\bar{1}$10]. 
The contrasts (colors, online) of the circles represent the spin polarization orientation of each state as defined in Fig. 4(b). 
The radii of the circles $R_{k_{\parallel}, E}$ are defined by the function,
\begin{eqnarray}
R_{k_{\parallel}, E}\propto \left|\sum_{i=1}^6\left(|\langle \phi_{CW}^i |\Psi_{k_{\parallel}, E}\rangle |^2 - |\langle \phi_{CCW}^i |\Psi_{k_{\parallel}, E}\rangle |^2\right)\right|,
\end{eqnarray}
where $|\phi_{CW}^i\rangle$ ($|\phi_{CCW}^i \rangle$) represents the atomic orbital in the $i$ th Sn layer with CW (CCW) spin polarization, 
and $|\Psi_{k_{\parallel}, E}\rangle$ is the eigenfunction of the calculated state at ($k_{\parallel}$, $E$). 
Thus, the large circles in Fig. 4(a) represent the states which are spin-polarized towards CW or CCW directions and localized in the surface Sn layers. 

As shown in Fig. 4(a), the calculated spin-polarized upper cone $S_1$ and lower cone $S_2$ agree with those experimentally observed.
All states are calculated with smaller binding energies of $\sim$80 meV than what are observed.
Since the films we measured are double-domain ones, the overlap of the calculated states along [110] and [$\bar{1}$10] should be observed by ARPES.
$S_1$ is isotropic along both lines and the observed $S_2$ bands would originate from [$\bar{1}$10].
In addition, along [110], there is another band $S_2'$ which agrees with that observed along $k_x$ on the 34 ML film (see Fig. 2(j)).
The upper edge of the heavy-hole-like QW states is also spin polarized around $\bar{\Gamma}$, possibly due to the hybridization with DC.
It could be the origin of the circular dichroism observed at $|k_y| >$0.15 \AA$^{-1}$.
The bandgap between bulk-like QWs around $\bar{\Gamma}$ is 380 meV.
Counting in whole SBZ, the gap is 230 meV: the conduction band minimum is around $\bar{J}'$.
For thinner films, the agreement between experiment and calculation is rather poor since the substrate effect cannot be neglected.

In order to understand the role of surface Bi, we calculated Sn slabs with various surface structures without Bi.
The slabs with dangling bonds (such as the (2$\times$1) clean surface) made the other surface states dispersing around $E_{\rm F}$ which is not observed in this work.
On the other hand, once the dangling bonds are saturated by adatoms, such as Bi and H, these artificial surface states disappeared.
Therefore, the main role of Bi on the Bi/Sn(001) films is probably to saturate the surface dangling bonds.
The detailed results are in the supplemental material \cite{SM}. 

The spin polarization orientations calculated for $S_1$ (CW) and $S_2$ (CCW) are opposite to those observed by spin-resolved ARPES: CW (CCW) for $S_2$ ($S_1$).
This is due to the photoelectron spin-flipping due to the polarization of photons \cite{Jozwiak13}.
According to ref. \cite{Jozwiak13}, the linearly-polarized photons in our experimental geometry flip the spin polarization of the photoelectrons measured along [$\bar{1}$10] with respect to the initial states.

In order to explain the surface-state band evolutions observed in this work, we propose the following scenario (see Fig. 4 (d)).
(i) For very thin films (12 ML), the strong interference between the top and bottom surfaces opens the bandgap on DC \cite{Shan10, Zhang10, Sakamoto10}.
(ii) By increasing the thickness, the topologically protected DC appears in the bandgap because the interference between both faces become weak. It corresponds to thicknesses from 20 to 24 MLs in this work.
(iii) At higher thicknesses, QW states with heavy-hole character appears in the gap and hybridize with DC. In this case, the pristine Dirac point lies below the topmost heavy-hole-QW state. Note that the upper part of the DC is still dispersing continuously in the gap between bulk-like QW states. This case corresponds to the films with thicknesses above 30 ML and consistent with what is calculated.
(iv) At infinite thickness, the gap between heavy-hole and light-hole QWs closes and the system becomes a 3D zero-gap semimetal.
Based on this model, phase (ii) and (iii) can be regarded as TI with spin-polarized DC.
Since such TI phase can be achieved only for finite thicknesses, it could be categorized as ``quasi-3D" TI. 
Moreover, since the QW states in (ii) and (iii) have the inverted band structure which is necessary for TI, it would also hold the 1D edge states at the edge of the QW, as the case of HgTe QW \cite{Konig07, Xu13}.

In conclusion, we have reported Dirac-cone-like dispersion of surface states of $\alpha$-Sn(001) films covered with 1 ML of Bi grown on InSb(001) substrates.
Both the spin-resolved ARPES and circular dichroism of ARPES indicate helical spin polarization of the Dirac-cone-like surface states.
A bandgap in the film is estimated to be 230 meV showing that a new type of TI phase can be fabricated with ultrathin films of zero-gap semiconductors. 
Based on the evolution of Dirac-cone-like surface states with film thicknesses, we have demonstrated a new opportunity to fabricate ultrathin TI films with thicknesses down to few nm on conventional semiconductors.
These results should offer new perspectives of applications in miniaturized electronic/spintronic devices.

\newpage

\section*{Supplementary material to: Dirac cone with helical spin polarization in ultrathin $\alpha$-Sn(001) films}

\subsection{Sample growth procedure}
The InSb(001) substrates were cleaned by repeated cycles of sputtering and annealing up to 700 K.
Prior to Sn deposition, the substrates were covered with 1 ML of Bi. 
We also grew the Sn films without pre-deposition of Bi for the sake of comparison.
Sn was deposited from a Knudsen cell and the as-grown films were annealed up to 470 K.
The deposition rate of Bi and Sn were monitored by a quartz microbalance and calibrated using both Auger and core-level photoelectron spectra.

\subsection{Characterization of Bi/Sn(001) films}
Figure 1S (a) shows the core-level spectra of the Sn(001) film.
Bi $5d$ is much more intense than Sb $4d$ and In $4d$, implying that Bi segregates at the surface during the Sn growth and forms the topmost atomic plane of the sample.
Figure 1S (b) shows Sn 4$d$ core levels from 24 ML Bi/Sn film (the upper spectrum) and that grown on the sample holder (the lower one), which is made of polycrystalline Ta plate.
Since the $\beta$ phase is stable at room temperature, Sn on the Ta plate must be $\beta$-Sn.
The spectrum from the Bi/Sn(001) film is fitted well with Sn 4$d$ (dashed line) and Bi 5$d$ (gray area) components.
In contrast, on Ta, an additional broad component (light blue area) is required to fit the spectrum.
It would originate from contaminants (possibly, oxide layers on Ta) remained on the Ta plate: the cleaning procedure used for InSb(001) is not efficient for Ta.
The Sn 4$d$ peaks from the Bi/Sn film are shifted with 230 meV from the $\beta$ Sn on the Ta plate.
This core-level shift is consistent with the previous report about $\alpha$-Sn(111) grown on InSb(111) \cite{Hernandez85}.

Figures 1S (c) and (d) are the low-energy electron diffraction (LEED) patterns of the InSb(001) clean surface and the Bi/Sn(001) film, respectively.
A low-background and sharp-spot LEED pattern in Fig. 1S (d) indicates the growth of  a well-ordered $\alpha$-Sn(001) film with double-domain (2$\times$1) fractional spots.
As shown by the dashed squares in Fig. 1S (c) and (d), there are almost no difference of lattice constants between the InSb(001) clean surface and the Bi/Sn(001) film.
It indicates the in-plane lattice constant of the Bi/Sn(001) film to be close to that of InSb (4.58 \AA).
This is consistent with that of $\alpha$-Sn (4.59 \AA) but quite different from those of $\beta$-Sn ($a$ = 5.83 \AA\ and $c$ = 3.18 \AA).
This $\alpha$-Sn film is stable even after post annealing above room temperature.
The Sn film without Bi also showed the double-domain (2$\times$1) LEED pattern (not shown).

A (2$\times$1) pattern suggests a dimer-row structure as in Si(001) and Ge(001) surfaces \cite{Schlier59}.
Since the films show double-domain patterns, there are variations in the film thickness of more than 1 ML: 
the dimer-row rotates from 90$^\circ$ at each atomic step.
In the Bi/Sn(001) films, the topmost layer consists of Bi, as indicated by the core-level spectra.
Based on the (2$\times$1) periodicity, the topmost Sn dimers are possibly replaced with Bi, as modeled in Fig. 4 (c) in the main text.
In this model, surface dangling bonds of the surface Sn atoms are saturated by Bi.

\subsubsection{In interdiffusion}
From the In-enriched clean surface of InSb, In interdiffuses in the Sn overlayer \cite{Fantini00, Betti02, Magnano02}. 
It is reported that the maximum thickness of $\alpha$-Sn(111) films grown on Sb-enriched InSb(111) substrates is much larger than that grown on the substrates prepared by the cycles of sputtering and annealing, whose surface would be the In-enriched \cite{Kasukabe88, Osaka94}.
It suggests that the In interdiffusion into upper Sn layers reduces the quality of the epitaxial Sn film.
We deposited 1 ML of Bi prior to the Sn deposition, expecting that the initial Bi layer may act as a barrier of such In interdiffusion.

Fig. 2S is the In 4$d$ levels after Sn growths with/without Bi. It shows two In $4d$ components with the peak positions of 4$d_{3/2}$ at 58.1 ($A$) and 58.6 eV ($B$), respectively.
Both of them were observed in the previous study \cite{Magnano02}.
In ref. \cite{Magnano02}, the authors reported that (i) both components increase with the substrate temperature during the Sn growth and (ii) $B$ corresponds to the In atoms that are chemically bonded with Sn.
Based on ref. \cite{Magnano02}, Fig. 2S shows that sizable interdiffusion of In occurs in both films, indicating that In interdiffuses into the Sn layer irrespective of the existence of pre-deposited Bi.
However, we could observe sharp LEED patterns in both cases.
It suggests that the In interdiffusion into Sn layers is not so serious problem for the $\alpha$-Sn growth on InSb(001), other than the case of the InSb(111) surfaces.

\subsection{The role of Bi onto the surface electronic structure}
\subsubsection{ARPES results}

Figure 3S shows the ARPES intensity plots of $\alpha$-Sn(001) films grown with (a, b) and without (c, d) Bi at rather thin thicknesses.
In these thicknesses, the Bi-covered film shows surface bands $S_1$ and $S_2$ and the dispersion of $S_2$ is isotropic. 
In contrast, the band dispersion of the ``bare" film shown in Fig. 3S(c, d) are not isotropic but clearly exhibits the measurement-plane dependence.
With the thickness around 10 ML, there is no QW states around $E_{\rm F}$, which are possibly the origin of the anisotropic surface-band dispersions of thick Bi/Sn(001) films.
Therefore, the ARPES results of the ``bare" film imply the existence of the surface atomic structure forming the surface states that is absent on the Bi/Sn(001) films.

\subsubsection{Theoretical calculations}

Figure 4S (a) is the same figure as Fig. 4 (a) in the main text.
Figure 4S (b) is calculated from the same slab as (a), but the atomic orbitals of surface Bi are included to calculate the radii of the circles for each electronic state.
As shown in Figs. 4S (a) and (b), there is almost no change on the spin-polarized surface states, indicating that the Dirac cone observed/calculated in this work is not the surface states localized at the surface Bi atoms. 

Figure 4S (c-e) are the calculated band structures of the slabs with different surface atomic structures.
Figure 4S (c) is obtained from the Bi/Sn(001)-(2$\times$1) surface, the same atomic structure as Figs. 4S (a, b).
There is no surface bands dispersing around the Fermi level in the projected bulk band gap, except for the Dirac cone at $\bar{\Gamma}$.
The clean Sn(001) surface is modeled by replacing the surface Bi of Fig. 4 (c) in the main text with Sn, resulting the metallic surface states dispersing across the Fermi level, as shown in Fig. 4S (d).
These surface states originate from the dangling bonds of the topmost Sn atoms.
The dangling bonds can be saturated by adatoms, $e.g.$ hydrogen.
This saturation moves the metallic surface states away from the Fermi level, as shown in Fig. 4S (e).
On such slabs with dangling-bond saturation, the only surface states around the Fermi level is Dirac-cone-like spin-polarized ones.
These results indicate that the saturation of dangling bonds on the surface is the essential condition of the surface electronic structure with single Dirac cone.
Therefore, while the Bi dimer-row model proposed in this work is just an assumption, the surface Bi probably saturates the surface dangling bonds on the Sn(001) surface somehow.

At $\bar{\Gamma}$ in Figs. 4S (d) and (e), there is a gap of 225 and 170 meV, respectively. 
It would be due to the hybridization between the Dirac cones localized at the top and the bottom layers. \cite{Zhang10, Luo10}.
Strictly speaking, there is also a small gap on the Bi/Sn(001) slab with 35 meV.
The difference of the gap sizes would be due to the different degree of the top-bottom interference influenced by each surface structure.
This finite gap on the surface states was not observed experimentally in this work. 
Probably, it is because the actual films are asymmetric with different structures at the top and the bottom faces, in contrast to the symmetric slabs we adopted for theoretical calculations.
This asymmetry can protect the Dirac cone from the interference between the top and bottom surface states, as observed on Sb(111) thin films \cite{Bian12}.

\subsection{Initial-state parities with respect to the measurement plane}

The experimental geometry of the ARPES experiments with linearly-polarized photons influences the observed states, because of the parities of the initial states with respect to the measurement planes.
In our experimental condition, the photon-incident plane is ($\bar{1}$10) and the electric field of linearly-polarized photons lies in the incident plane.
In this condition, the dipole selection rule allows the photoelectron excitations from initial states with odd (even) parities with respect to the measurement plane along [$\bar{1}$10] ([110]).
Figures 5S (c) and (d) show the calculated states localized in subsurface Sn layers as Fig. 4S (a) but the radii of the circles represents the atomic-orbital contributions with odd and even parities with respect to the planes in which each $k$ vector lies, respectively.
Indeed, ARPES data shown in Fig. 5S (a) and (b) are well reproduced in Fig. 5S (c) and (d), respectively.
In addition, both initial states corresponds with $S_2'$ and heavy-hole like QWs, that are observed in Fig. 5S(b) (along [110]), has a large contribution from even states (Fig. 5S (d)).

\begin{figure}[p]
\includegraphics[width=80mm]{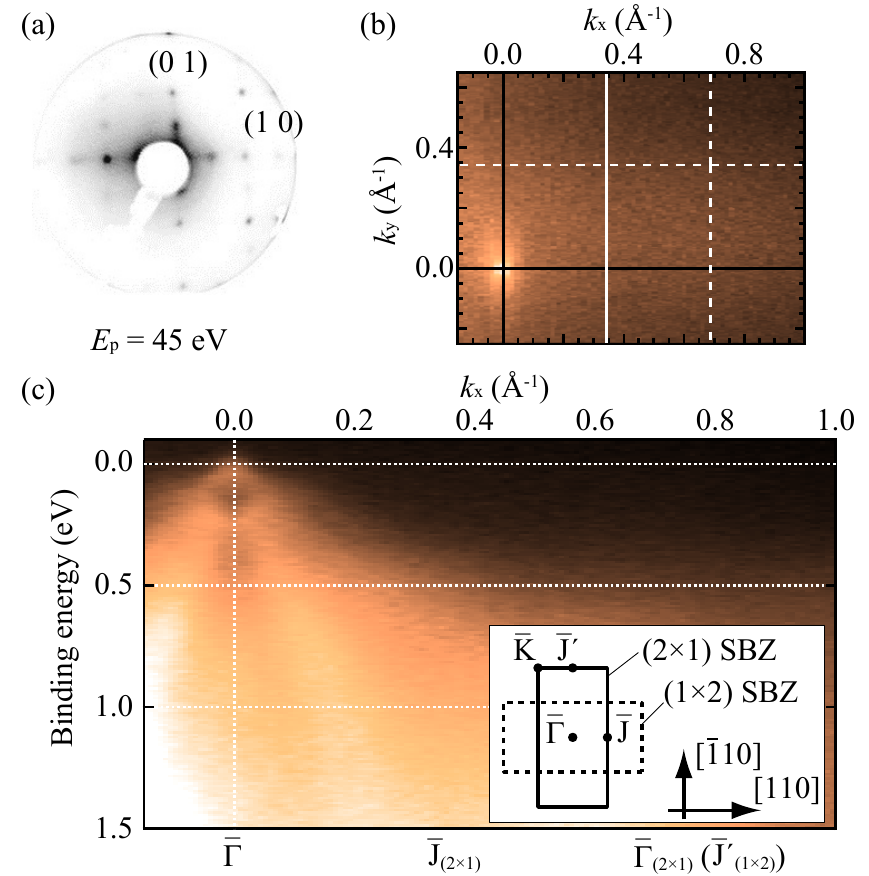}
\caption{\label{fig1} (Color online).
(a) LEED pattern of the 12 ML Sn(001) film.
(b, c) Fermi contour (b) and the band dispersion along [110] (c) measured by ARPES with h$\nu$ = 19 eV.
Solid and dashed lines in (b) correspond to (2$\times$1) and (1$\times$2) surface Brillouin zones depicted in the inset in (c), respectively.
The inset also shows our definition of the coordinates.
$k_x$ ($k_y$) is defined as parallel to [110] ([$\bar{1}$10]).
All data were taken at room temperature (RT).
}
\end{figure}

\begin{figure*}[p]
\includegraphics[width=150mm]{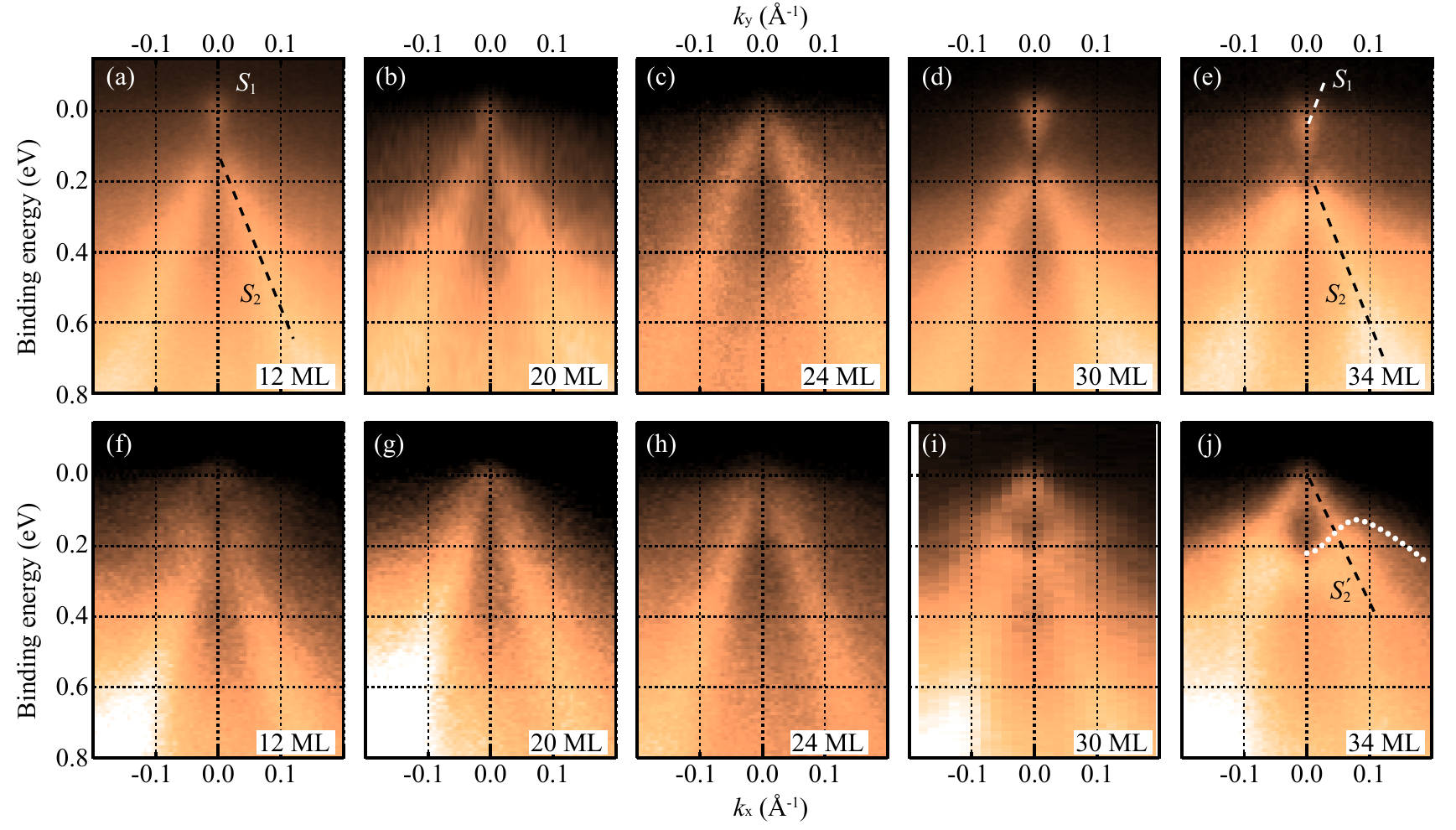}
\caption{\label{fig2} (Color online).
(a-e) ARPES intensity plots of 12-34 ML films along [$\bar{1}$10] measured with h$\nu$ = 19 eV at RT.
(f-j) The same as (a-e) but taken along [110].
Llines are guides for eye.
}
\end{figure*}

\begin{figure}[p]
\includegraphics[width=80mm]{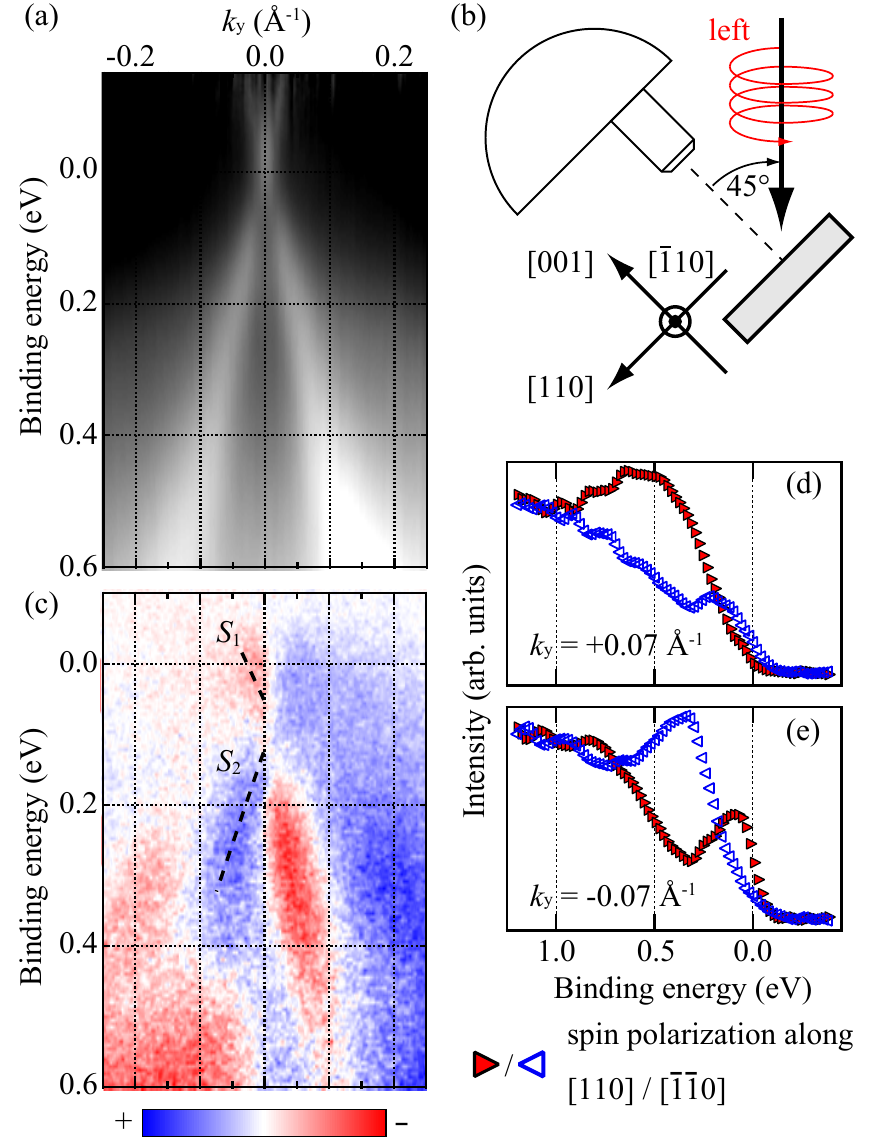}
\caption{\label{fig3} (Color online). 
(a) ARPES intensity plot of the 24 ML film along [$\bar{1}$10] measured with h$\nu$ = 19 eV, divided by the Fermi-Dirac distribution at 450 K function convolved with the instrumental resolution.
(b) Experimental geometry in this work. 
(c) ARPES circular-dichroism plot of the 30 ML film taken with 19 eV photons at RT. Dichroism is obtained by subtracting the intensities of left-circularly polarized photons from those of right-circularly polarized ones.
(d, e) Spin-resolved ARPES spectra taken at RT with spin polarization toward [110]/[$\bar{1}\bar{1}$0]. The acceptance angle for the spin-resolved ARPES corresponds to $\Delta k_y = \pm$0.06 \AA$^{-1}$.
}
\end{figure}

\begin{figure}[p]
\includegraphics[width=80mm]{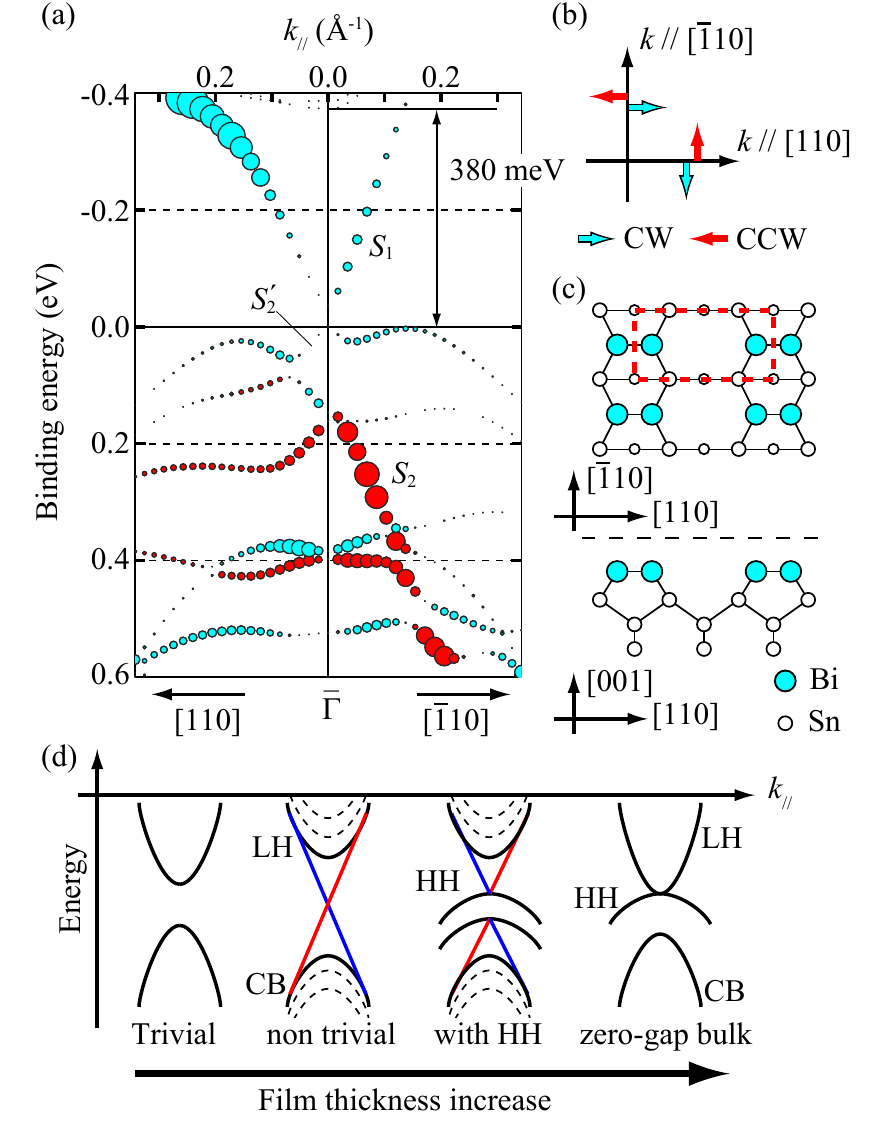}
\caption{\label{fig4} (Color online). 
(a) Calculated band structure along [110] and [$\bar{1}$10] for the slab of 32 Sn layers covered with Bi dimers. The radii of the circles are defined by the function in the text (1).
The contrasts (colors, online) of the circles represent the spin polarization of each state.
(b) Definition of the spin polarization directions.
(c) Schematic picture of the assumed atomic structure for the DFT calculation.
A dashed rectangle denotes the (2$\times$1) unit cell.
(d) Schematic drawings of the surface-state evolution on Sn(001) films with respect to light-hole (LH), heavy-hole (HH), and conduction-band (CB) bulk-like QW states.
}
\end{figure}

\renewcommand{\thefigure}{\arabic{figure}S}
\setcounter{figure}{0}

\begin{figure}[p]
\includegraphics[width=80mm]{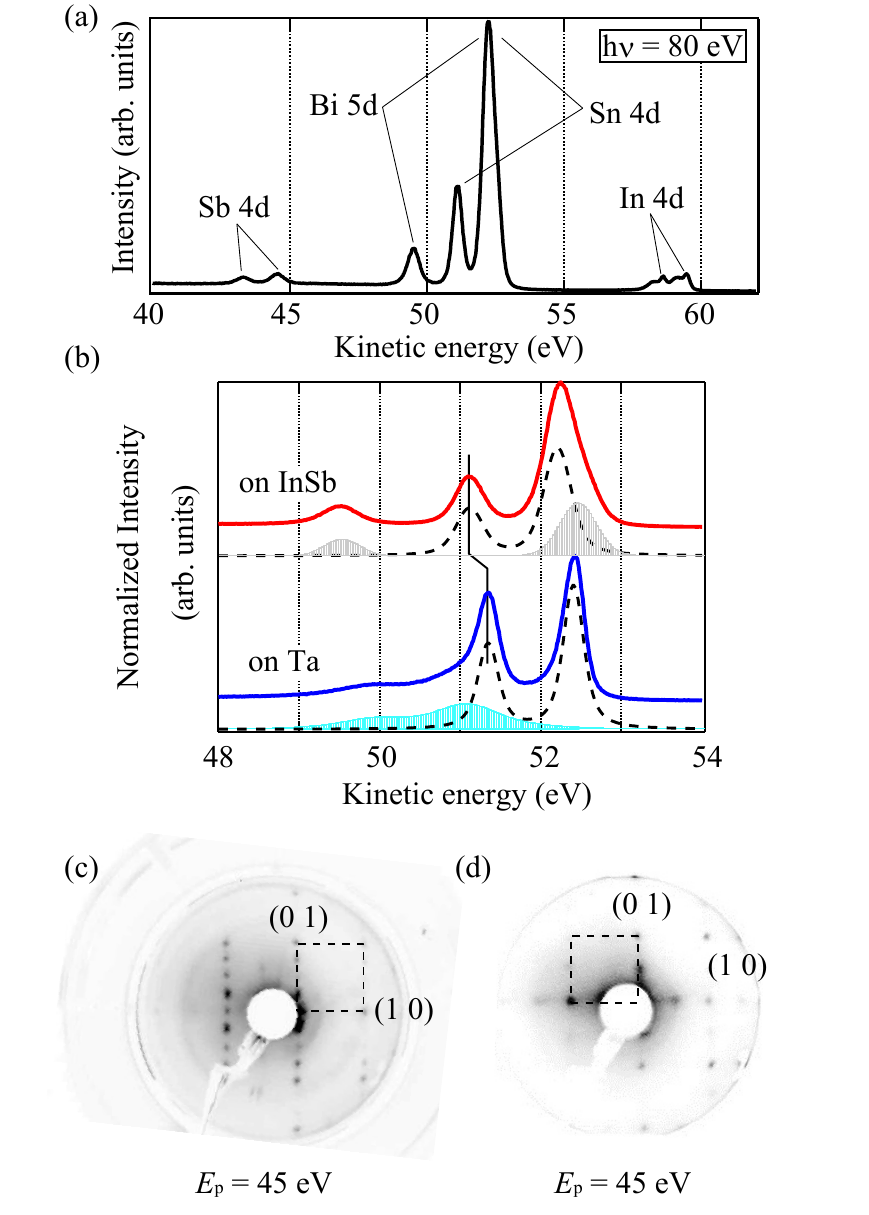}
\caption{\label{fig1}
(a) Core-level spectrum for the 24 ML Bi/Sn(001) film grown on InSb(001).
(b) Core-level spectra for the 24 ML Bi/Sn(001) film grown on InSb(001) and $\beta$-Sn film grown on the sample holder (Ta plate).
(c, d) LEED pattern of (c) InSb(001)-c(2$\times$8) (clean surface) and (d) the 12 ML Bi/Sn(001) film.
Dashed squares in (c) and (d) are the same size as a guide to the eye.
All data were taken at RT.
}
\end{figure}

\begin{figure}[p]
\includegraphics[width=80mm]{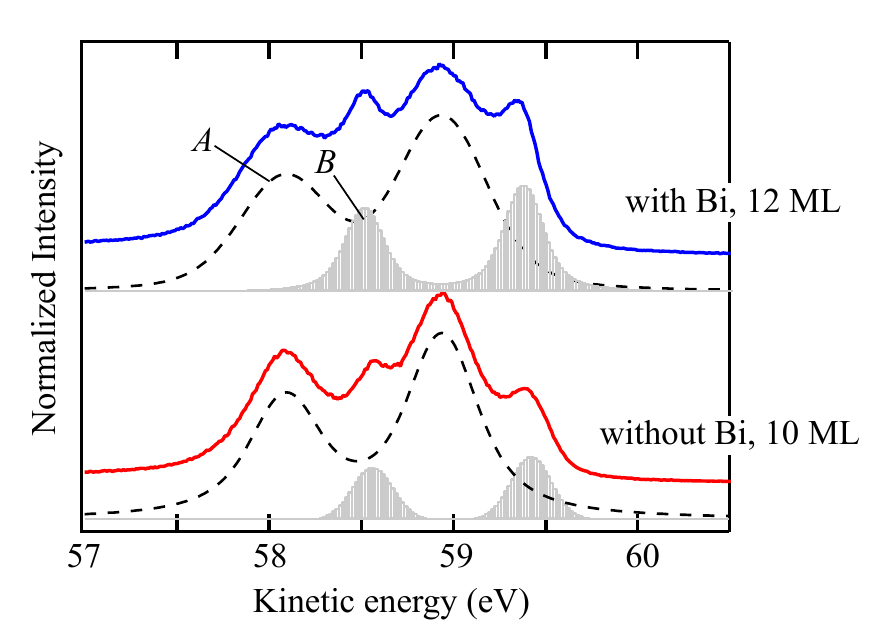}
\caption{\label{fig2}
Indium $4d$ Core-level spectra for Sn(001) films with and without Bi taken at RT with h$\nu$ = 80 eV.
}
\end{figure}

\begin{figure}[p]
\includegraphics[width=80mm]{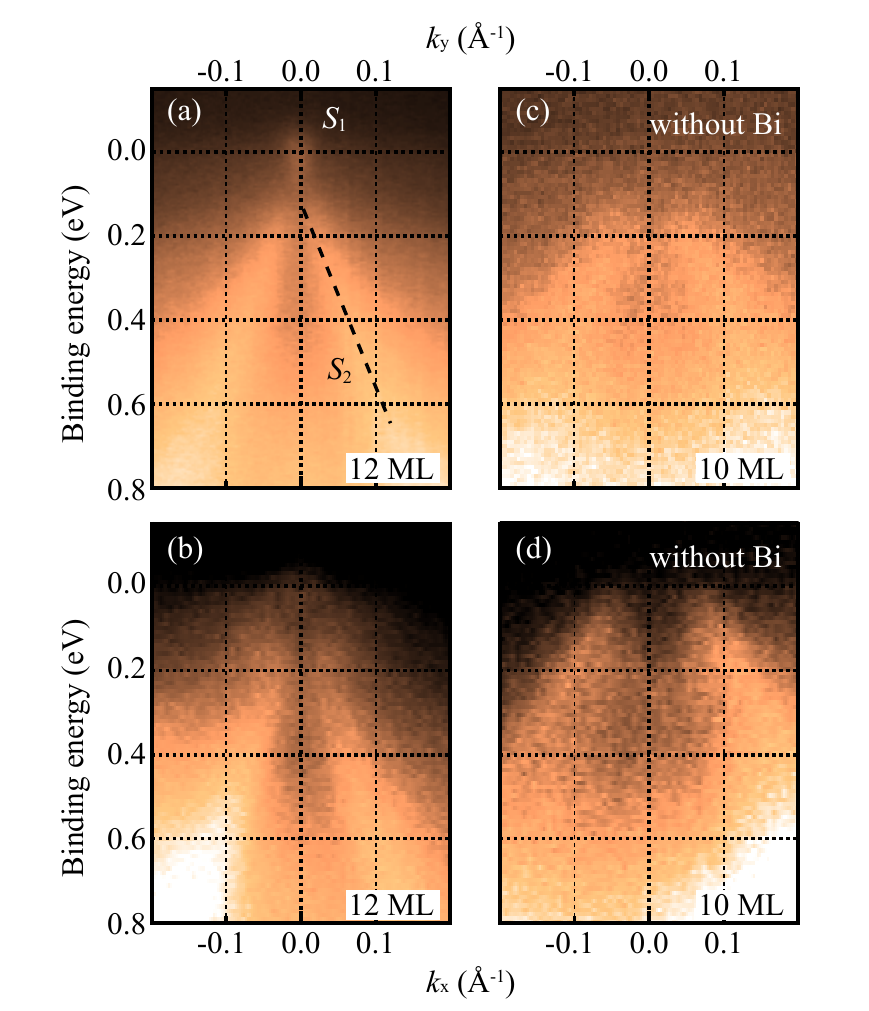}
\caption{\label{fig3}
ARPES intensity plots of Sn(001) films grown with (a, b) and without (c, d) Bi.
The definition of the coordinates are the same as Figures 1 and 2 in the main text.
(a) and (b) are copied from Fig. 2 (a) and (f) in the main text.
}
\end{figure}

\begin{figure}[p]
\includegraphics[width=80mm]{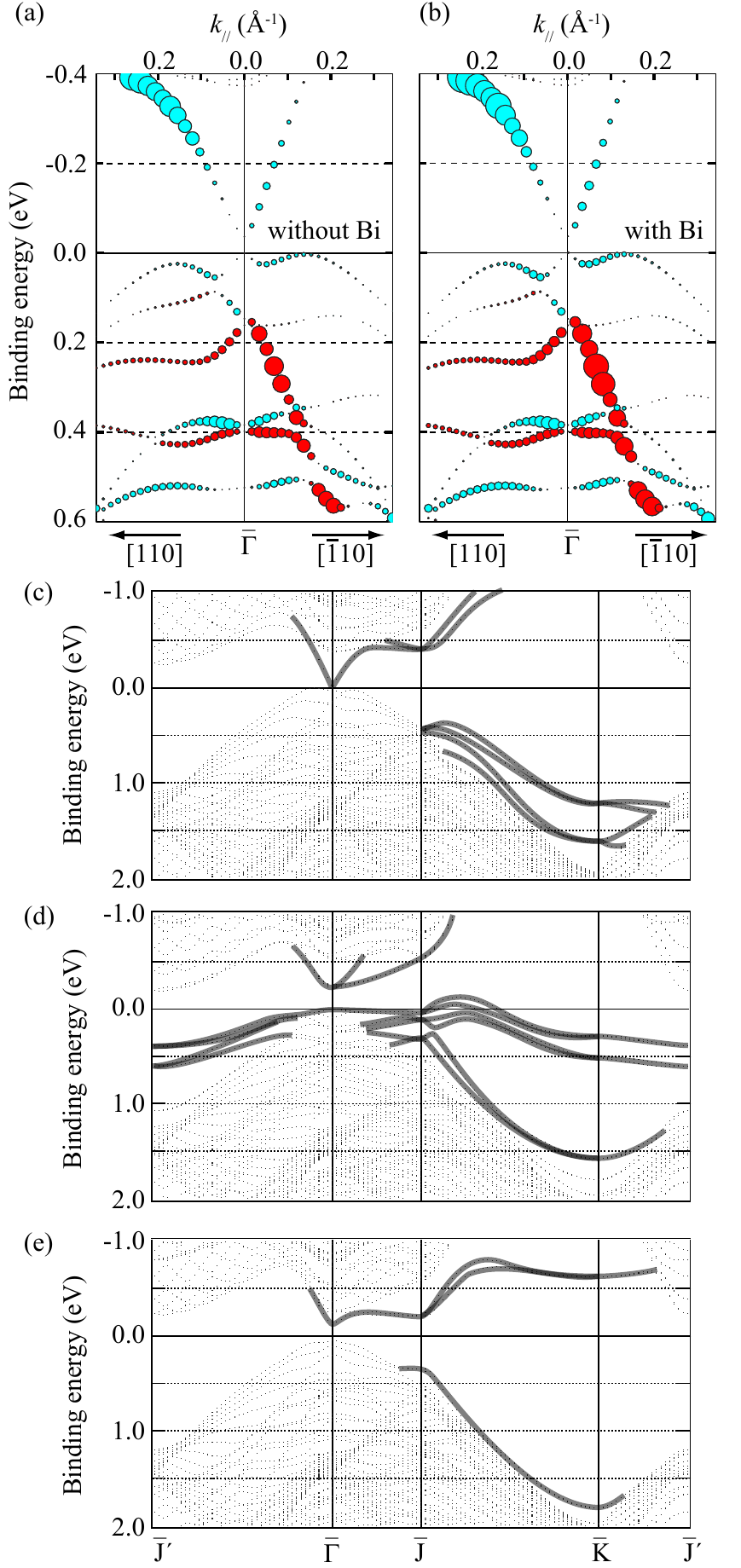}
\caption{\label{fig4}
(a, b) Calculated band structures for the Bi/Sn(001)-(2$\times$1) slab. (a) The same as Fig. 4(a) in the main text. 
(b) the same as (a) but the radii of the circles are obtained including the contribution from the atomic orbitals of surface Bi.
(c) Calculated band structure for the Bi/Sn slab with whole surface Brillouin zone (SBZ). Solid lines represent surface states.
(d, e) The same as (c) but calculated with the slabs whose surfaces are modeled by (d) the Sn(001)-(2$\times$1) dimer row with bare dangling bonds and (e) the H/Sn(001)-(2$\times$1) dimer row with all the dangling bonds saturated by hydrogen.
}
\end{figure}

\begin{figure}[p]
\includegraphics[width=80mm]{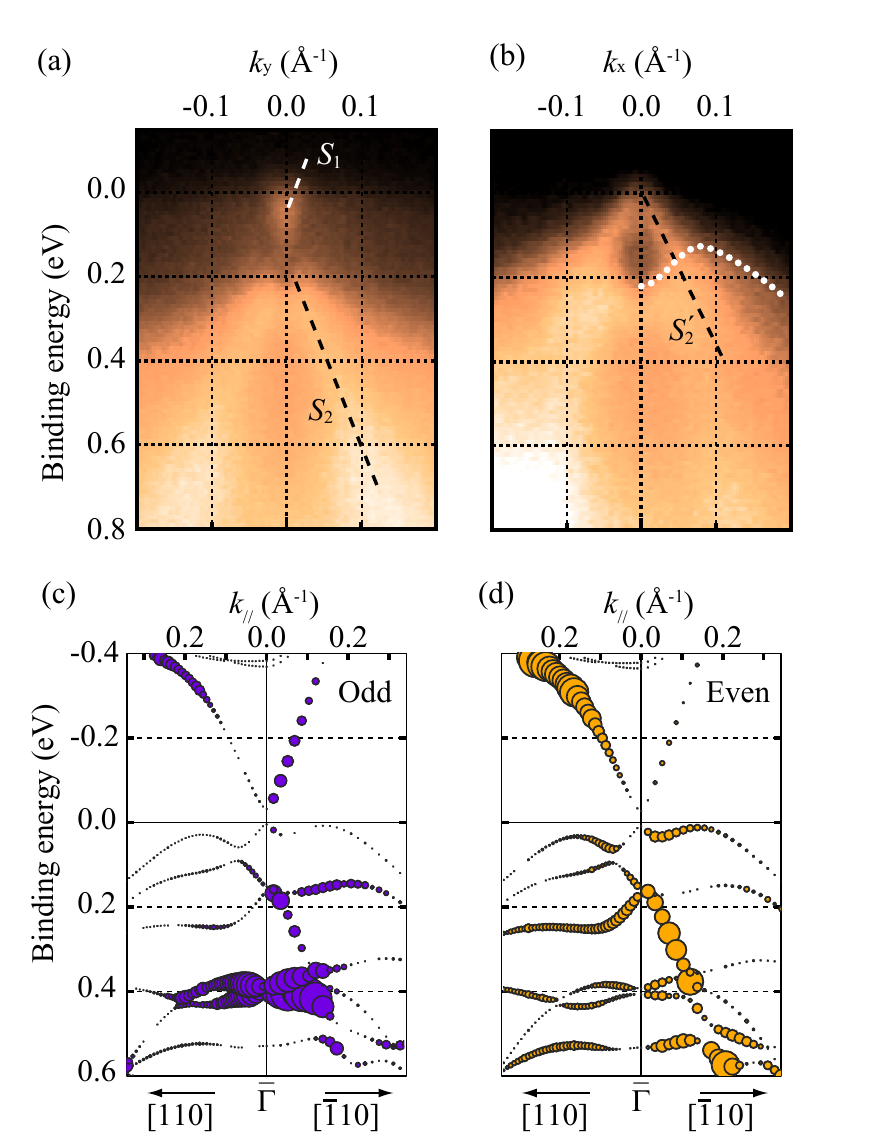}
\caption{\label{fig5}
ARPES intensity plots of the 34 ML Bi/Sn(001) film along [$\bar{1}$10] (a) and [110] (b), copied from Fig. 2(e) and (j), respectively.
(c) Calculated band structure corresponding to the ARPES result shown in (a).
(d) The same as (c) but corresponding to the result in (b).
}
\end{figure}

\end{document}